\documentclass[11pt,draftclsnofoot,onecolumn]{IEEEtran}

\usepackage{graphicx}
\usepackage{epsfig}
\usepackage{cite}
\usepackage{amsmath,amsfonts,amssymb,amsthm,mathrsfs}
\usepackage{algorithm, algorithmic,centernot}
\usepackage{tikz}
\usepackage{etoolbox}
\makeatletter
\patchcmd{\@makecaption}
  {\scshape}
  {}
  {}
  {}
\makeatletter
\patchcmd{\@makecaption}
  {\\}
  {.\ }
  {}
  {}
\makeatother

\theoremstyle{plain}
\newtheorem{thm}{Theorem}
\newtheorem*{thm*}{Theorem}

\newtheorem{lem}{Lemma}
\newtheorem*{lem*}{Lemma}
\newtheorem{cor}{Corollary}

\newtheorem*{prop*}{Proposition}

\theoremstyle{definition}

\begin{document}
\title{ Entropy rates for Horton self-similar trees}
\author{Evgenia V. Chunikhina\footnote{
School of EECS, Oregon State University, Corvallis, OR 977330
\texttt{chunikhe@oregonstate.edu}}}
\date{ }
\maketitle

\abstract
In this paper we examine planted binary plane trees. First, we provide an exact formula for the number of planted binary trees with given Horton-Strahler orders. Then, using the notion of entropy, we examine the structural complexity of random planted binary trees with $N$ vertices. Finally, we quantify the complexity of the tree's structural properties as tree grows in size, by evaluating the entropy rate for planted binary plane trees with $N$ vertices and for planted binary plane trees that satisfy Horton Law with Horton exponent $R$.


\section{Introduction}
\label{sec:introduction}

Tree-like structures are among the most widely observed natural patterns, occurring in the applied fields of study as diverse as river and drainage networks, botanical trees and leaves, blood systems, crystals, and lightening.
In addition, many processes like branching processes, percolation, nearest-neighbor clustering, binary search trees in computer science, spread of a disease, spread of news on social platforms, or propagation of gene traits can be represented as trees; see \cite{turcotte1999inverse,sorting1973searching,yakovlev2005inverse,zaliapin2005inverse,zaliapin2006hierarchical,zaliapin2012tokunaga,horton1945erosional,strahler1957quantitative,yekutieli1994horton,kemp1979average,peckham1995new,devroye1994note,newman1997fractal, turcotte1998networks,veitzer2000random, burd2000self,dodds2000scaling} and references therein.

An important measure of branching complexity of tree graphs was proposed in hydrology by Horton \cite{horton1945erosional} and Strahler \cite{strahler1952hypsometric,strahler1957quantitative}. The Horton-Strahler ordering scheme assigns an order to each tree branch in accordance with its hierarchial importance. This measure found its practical application in many different areas, ranging from hydrology and biology to computer science and neuroscience \cite{burd2000self,devroye1994note,dodds2000scaling,kemp1979average,peckham1995new,turcotte1998networks,veitzer2000random,yekutieli1994horton}.
In particular, Horton self-similarity is an important property, that describes the geometric decrease of Horton-Strahler numbers. In recent years, the questions related to Horton self-similarity were addressed in a variety of scientific publications \cite{milne2017horton,neveu1986erasing,dodds1999unified,peckham1995new,dodds2000scaling,tarboton1988fractal,gabrielov1999exactly,burd2000self,zaliapin2012tokunaga,kovchegov2017horton}.
In this work, we consider a space of uniformly distributed planar planted binary trees and determine the number of trees with given structural features, such as the number of vertices in a tree, the Horton-Strahler order of a tree, and the Horton-Strahler numbers. Also we use entropy to study the structural complexity of uniformly distributed planted binary plane trees with given Horton-Strahler numbers. Furthermore, we find closed-form formula for the entropy rate, that describes the growth of the entropy as the number of vertices of the tree grows to infinity. In particular, we consider a special class of binary trees that satisfy Horton law with a given Horton exponent $R$ and find a closed-form formula for its entropy rate. Note that the uniform distribution on the space of planted binary plane trees with $N$ vertices is different from the uniform distribution over the space of planted binary non-plane trees, induced by the critical binary Galton-Watson process, conditioned on having $N$ vertices.

The paper is organized as follows. In Section \ref{sec:preliminaries}, we provide main definitions and notations that are used throughout this paper. The formula for the number of planted binary trees with specific Horton-Strahler numbers is given in Section \ref{sec:main-formula}. The main results are provided in Section \ref{sec:entropy-catalan}, where, using the notions of entropy and entropy rate, we quantify the structural complexity of Horton self-similar trees and growing tree models. All proofs are provided in the Appendix \ref{sec:Appendix}.

\section{Preliminaries}
\label{sec:preliminaries}

\subsection{Planted binary plane trees}
\label{ssec:planted_trees}

In this paper, a $\mathit{tree}$ is defined as an acyclic connected graph. A tree with one vertex labeled as the $\mathit{root}$ is called a $\mathit{rooted}$ $\mathit{tree}$. Presence of the root in a tree provides a natural child-parent relation between the neighboring vertices. More precisely, the $\mathit{parent}$ of a vertex is the vertex connected to it on the path down, towards the root and the $\mathit{child}$ of a vertex is the vertex connected to it on the path up, away from the root. Note that a vertex can have more than one child and every vertex except the root has a unique parent and a unique $\mathit{parental}$ $\mathit{edge}$ that connects a vertex to its parent. A $\mathit{leaf}$ is a vertex with no children. The $\mathit{degree}$ of a vertex is the number of edges incident to a vertex.

\begin{figure}
  \begin{center}
  \includegraphics[scale=0.3]{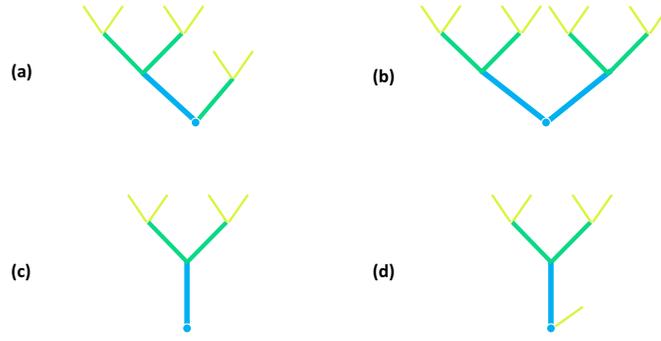}\\
  \end{center}
  \caption{An example of a full binary tree (a) and a perfect binary tree (b). The tree (c) is a planted binary plane tree. The root node is depicted at the bottom of the tree and has degree one. The tree (c) has $n = 4$ leaves, $2n = 8$ vertices, and $2n-1 = 7$ edges. The tree (d) is not a planted binary plane tree since its root node has degree two.}
  \label{planted_tree_example}
\end{figure}

A tree is called a $\mathit{binary}$ tree if each vertex has at most two children. A $\mathit{full}$ binary tree is a tree in which every vertex has either zero or two children. A $\mathit{perfect}$ binary tree is a binary tree in which all interior vertices have two children and all leaves have the same depth. In Figure \ref{planted_tree_example} (a) and (b) we depict full and perfect binary trees, respectively. A $\mathit{plane}$ tree is a rooted tree with a specified ordering for the children of each vertex. This ordering is equivalent to an embedding of the tree in the plane and provides a natural left and right orientations for the children. A $\mathit{planted}$ $\mathit{binary}$ $\mathit{plane}$ $\mathit{tree}$ is a rooted tree such that its root has degree one and every other vertex is either a leaf or an $\mathit{internal}$ $\mathit{vertex}$ of degree three (see section 7.2 in \cite{pitman2006combinatorial}). We denote a $\mathit{stem}$ to be the unique edge that connects the root vertex with its only child. Assuming the tree grows from the root vertex upwards, the root vertex is located at the bottom of the stem. Every planted binary plane tree with $n$ leaves has $2n-1$ edges and even number of vertices $2n$, such that $n-1$ of them are internal. For examples, see Figure \ref{planted_tree_example} (c) and (d).

In this paper, we consider the space of finite unlabeled rooted binary plane trees with no edge length. We denote this space by $\mathcal{T}$. Let $\mathscr{T}_N \subset \mathcal{T}$ be the space of all planted binary plane trees with $n$ leaves and $N=2n$ vertices. The number of possible configurations of a planted binary plane tree with $N$ vertices is given by the $(n-1)$th Catalan number $\mathcal{C}_{n-1}$ \cite{pitman2006combinatorial} as follows
\begin{equation*}
\label{catalan_n_leaves}
|\mathscr{T}_N| = \mathcal{C}_{n-1} = \frac{1}{n}{2n-2 \choose n-1} = \frac{\left(2n-2\right)!}{n!\left(n-1\right)!},
\end{equation*}
where $n = \frac{N}{2}$ and   ${n \choose k} = \frac{n!}{k!(n-k)!}.$

\subsection{Horton-Strahler ordering}
\label{ssec:HS-ordering}

The Horton-Strahler ordering of the vertices and branches in a binary tree is performed, from the leaves to the root node, by hierarchical counting \cite{burd2000self,peckham1995new,horton1945erosional,strahler1957quantitative,newman1997fractal,kovchegov2016horton} as follows
 \begin{itemize}
   \item each leaf is assigned order $1$;
   \item an internal vertex with children of orders $i$ and $j$ is assigned the order $k = \max(i,j)+\delta_{ij},$ where $\delta_{ij}$ is the Kronecker's delta;
   \item the parental edge of the vertex has the same order as the vertex;
   \item a $\mathit{branch}$ of order $i$ is a sequence of neighboring vertices of order $i$ together with their corresponding parental edges.
 \end{itemize}

\begin{figure}
  \begin{center}
  \includegraphics[scale=0.4]{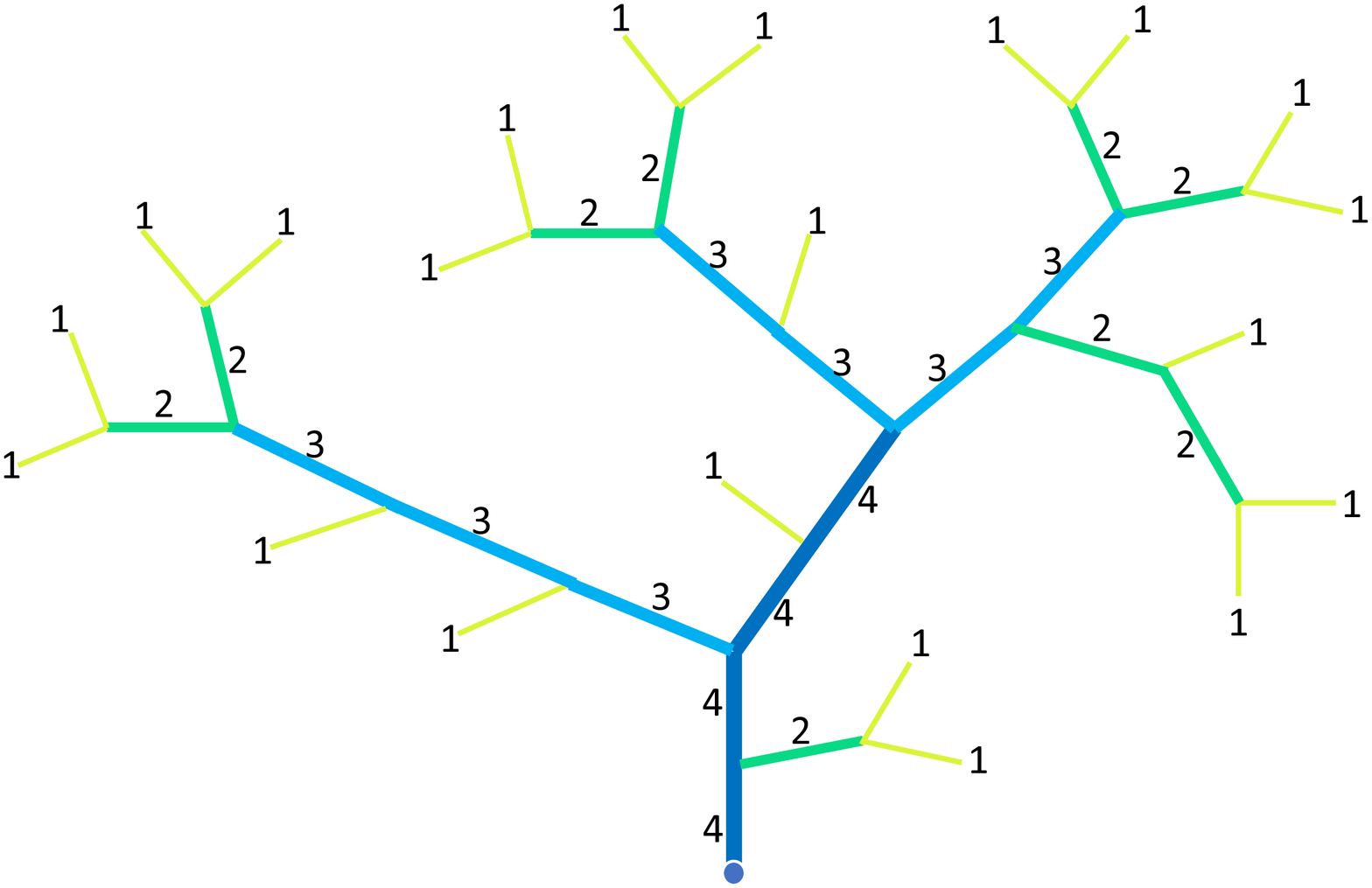}\\
  \end{center}
  \caption{Example of Horton-Strahler ordering of a planted binary plane tree. The tree has $N_4=1, N_3 = 3, N_2 = 8,$ and $N_1 = 21$. The number of vertices is $N = 2N_1 = 42$. Branches of order $4$ are depicted in indigo, branches of order $3$ in blue, branches of order $2$ in green, and branches of order $1$ in pear. The order of the tree is $K = 4$. The tree has only one branch of order $4$, although it consists of four edges. The stem of the tree is also of order $4$ and is a part of the branch of order $4$. The root node is depicted at the bottom of the tree and has order $4$.}
  \label{HS_example}
\end{figure}

The order $K$ of a non-empty tree is defined as the maximal order of its vertices. The $\mathit{Horton-Strahler}$ $\mathit{ordering}$ of a tree is defined as a set of numbers $N_i$, $i = \overline{1,K}$, where $N_i$ is the number of branches of order $i$. Note that
\begin{enumerate}
\label{cond1}
   \item in order to have a branch of order $i+1$ we need to have at least two branches of order $i$, i.e., $N_i\geq 2N_{i+1}$, $\forall i = \overline{1,K-1}$;
   \item a planted binary plane tree of order $K$ will have only one branch of order $K$, i.e., $N_K = 1$.
\end{enumerate}
In this work, we consider only {\it admissible sequences}, defined as the sequences $N_1, N_2, \cdots, N_K$ that satisfy the two conditions described above. We call an admissible sequence $N_1, N_2, \cdots, N_K$ a set of Horton-Strahler numbers.
To illustrate the Horton-Strahler ordering of a planted binary plane tree consider an example in Figure \ref{HS_example}.

Denote $\mathscr{T}_{N_1\cdots N_K}\subset \mathcal{T}$ be the space of all planted binary plane trees with Horton-Strahler numbers $N_1, N_2, \cdots, N_K$. Note that $\mathscr{T}_{N_1\cdots N_K} \subset \mathscr{T}_{N}$, where $N = 2N_1$. In the next section, we present our first statement that gives $|\mathscr{T}_{N_1\cdots N_K}|$ - the number of possible planted binary plane trees with Horton-Strahler numbers $N_1, N_2, \cdots, N_K$.

\section{Number of planted binary plane trees with given Horton-Strahler numbers}
\label{sec:main-formula}

\begin{lem}
\label{CountLemmaOne}
The number of planted binary plane trees of order $K$ with a particular set of Horton-Strahler numbers $N_1, N_2, \cdots, N_K$ and with $N = 2N_1$ vertices is given by the following formula
\begin{equation}
\label{main-formula}
|\mathscr{T}_{N_1\cdots N_K}| = 2^{N_1-1-\sum_{i=1}^{K-1}N_{i+1}}\prod_{i = 1}^{K-1} {N_i-2 \choose 2N_{i+1}-2} ,
\end{equation}
where ${n \choose k} = \frac{n!}{k!(n-k)!}.$
\end{lem}

The above lemma is known since its publication by Shreve in 1966 \cite{shreve1966statistical}. The detailed proof of Lemma \ref{CountLemmaOne} is provided in the Section \ref{ssec:proofCountLemmaOne}. Note that, although done for planted binary plane trees, the results of this lemma can be applied to the trees without a stem and to the trees with a $\mathit{ghost}$ $\mathit{edge}$ \cite{burd2000self,zaliapin2012tokunaga}.

\subsubsection{Example}
\label{sssec:example}

\begin{figure}
  \begin{center}
  \includegraphics[scale=0.5]{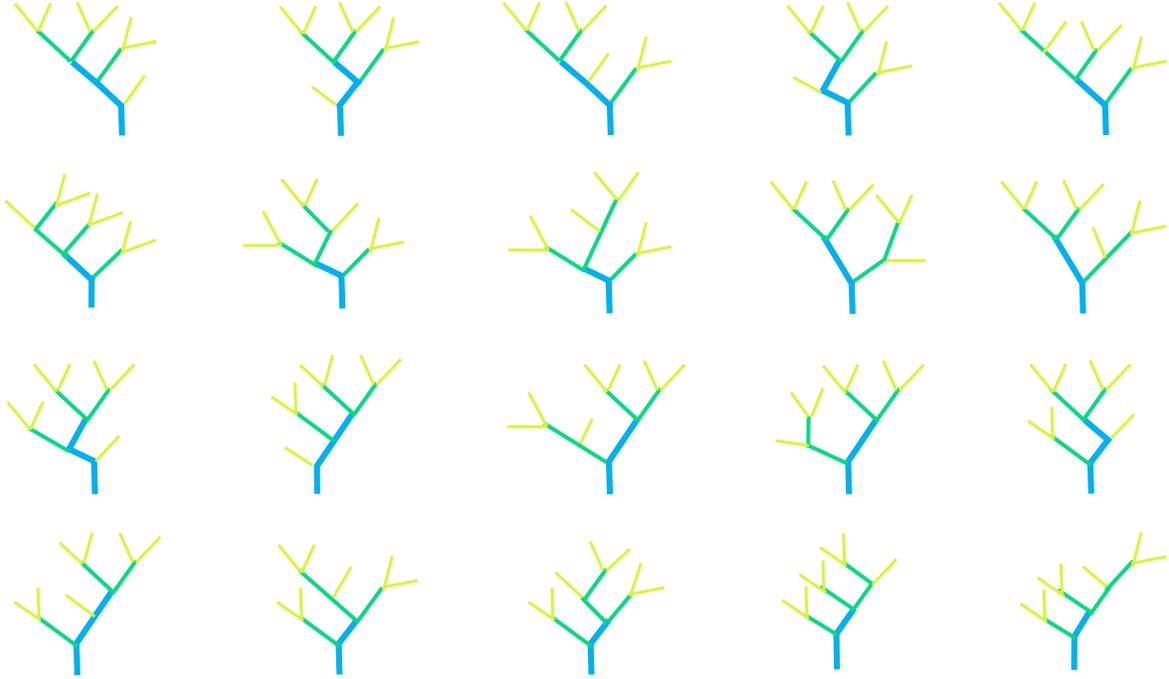}\\
  \end{center}
  \caption{An example of a space $\mathscr{T}_{7,3,1}$ of $20$ planted binary plane trees with $N_3=1, N_2 = 3$, $N_1 = 7$ and $N = 2N_1 = 14$ vertices. Branches of order $3$ are depicted in blue, branches of order $2$ in green, and branches of order $1$ in pear.}
  \label{20trees}
\end{figure}

Suppose we want to find the number of planted binary plane trees of order $K=3$ with the Horton-Strahler numbers $N_3 = 1, N_2 = 3$, $N_1 = 7$ and $N = 2N_1 = 14$ vertices. Using formula (\ref{main-formula}), we find that the number of such trees is
\begin{equation*}
\label{e7}
|\mathscr{T}_{7,3,1}| = 2^{N_1-1-\sum_{i=1}^{K-1}N_{i+1}}\prod_{i=1}^{K-1}{N_i-2 \choose 2N_{i+1}-2}= 2^{7-1-3-1}{7-2 \choose 6-2}{3-2 \choose 2-2} = 2^2\frac{5!}{4!1!}= 4\times 5 = 20.
\end{equation*}
In Figure \ref{20trees}, we depict all $20$ planted binary plane trees of order $3$ with $14$ vertices. In Table \ref{table1} we present the number of trees for different sets of Horton-Strahler numbers.

\begin{table}
\centering
\caption{Each entry in this table represents the number of trees for different sets of Horton-Strahler numbers: for the first two columns - $|\mathscr{T}_{N_1,N_2,N_3}|$ and for the second two columns - $|\mathscr{T}_{N_1,N_2,N_3,N_4}|$. The last row has the number of trees, when $N_1=30$.}
\begin{tabular}{c | c c c c}
  \hline
  $N_1$ & $N_2 = 2, N_3 = 1$ & $N_2 = 3, N_3 = 1$ & $N_2 = 4, N_3 = 2, N_4 = 1$ & $N_2 = 5, N_3 = 2, N_4 = 1$ \\
  \hline
  4 & 1 &  &  & \\
  5 & 6 &   &  & \\
  6 & 24 & 2 &  & \\
  7 & 80 & 20 &  & \\
  8 & 240 & 120 & 1 & \\
  9 & 672 & 560 & 14 & \\
  10 & 1792 & 2240 & 112 & 6\\
  11 & 4608 & 8064 & 672 & 108\\
  12 & 11520 & 26880 & 3360 & 1080\\
 \color{blue} 30 &\color{blue} 25,367,150,592 &\color{blue} 687,026,995,200 & \color{blue}1,580,162,088,960 & \color{blue}19,554,505,850,880\\
 \hline
\end{tabular}\\
\label{table1}
\end{table}

\section{Entropy and entropy rates}
\label{sec:entropy-catalan}

In this section, we use Shannon entropy to quantify the structural complexity of a tree. We propose an entropy based measure, namely the entropy rate of a tree, to examine how the structural complexity of a tree changes as a tree is allowed to grow in size. We find entropy rates for two types of trees: the planted binary plane trees with $N$ vertices and the planted binary plane trees with $N$ vertices that satisfy Horton law with Horton exponent $R$.

\subsection{Entropy and entropy rate for space $\mathscr{T}_N$}
\label{ssec:entropy_T_N}

Recall that entropy is a measure of the average uncertainty in the random variable \cite{cover2012elements,shannon1948mathematical}. For a discrete random variable $X$ with possible values $x_1, x_2, \cdots, x_n$ and probability mass function $P(X)$ the entropy of $X$ is defined by
\begin{equation*}
\label{entr_def}
H(X) = -\sum_{i=1}^{n}P(x_i)\log_2P(x_i) = -\mathbb{E}[\log_2P(X)],
\end{equation*}
where the quantity $0\log_20$ is taken to be $0$. The entropy is measured in bits. We can think of $\log_2P(x_i)$ as the uncertainty of the outcome $x_i$ (or the ``surprise" of observing $x_i$). Thus, the entropy is the average ``surprise". If possible values of $X$ are uniformly distributed, i.e., $P(x_i) = 1/n$, then there is a maximal uncertainty about the outcome, maximum ``surprise". In this case, the entropy achieves its maximal value $H(X) = \log_2n$. For example, the entropy of a fair coin toss is 1 bit. On the other hand, the occurrence of a certain event ($P(x_i)=1$, i.e., no ``surprise") has minimal uncertainty, which corresponds to the minimal value of entropy $H(X) = 0$.


From the information theory point of view, the entropy of a random variable can be thought of as an average number of bits required to describe the random variable \cite{cover2012elements}. Consider a random variable $X$ that has a uniform distribution
over $8$ outcomes, e.g., an eight-sided dice. The entropy of $X$ is $H(X) = -\sum_{i=1}^{8}\frac{1}{8}\log_2\frac{1}{8} = \log_28 = 3$ bits. A $3$-bit string takes on $8$ different values and is sufficient to describe $8$ outcomes of $X$. Note that all outcomes of $X$ have representations of the same $3$-bit length. Consider now a random variable $Y$ with a nonuniform distribution. Assume $Y$ can take $5$ possible values $\{y_1,y_2,y_3,y_4,y_5\}$ with corresponding probabilities $\left(\frac{1}{2},\frac{1}{8},\frac{1}{8},\frac{1}{8},\frac{1}{8}\right)$. The entropy of $Y$ is $H(Y) = -\frac{1}{2}\log_2\frac{1}{2}-4\frac{1}{8}\log_2\frac{1}{8} = 2$ bits.
Using Huffman coding technique we can encode outcome $y_1,y_2,y_3,y_4,y_5$ as strings $1,011,010,001,000$, respectively. Since we use shorter description for the more probable outcome $y_1$ and longer descriptions for the less probable outcomes $y_2,y_3,y_4,y_5$, the average description length is equal to the value of the entropy and is exactly $2$ bits.

Consider now a space $\mathscr{T}_N$, and let $P$ be a probability measure over $\mathscr{T}_N$. We define the entropy of a random planted binary plane tree $T_N \in \mathscr{T}_N$ as follows
\begin{equation*}
\label{entr_tree}
H(T_N) = -\mathbb{E}[\log_2P(T_N)],
\end{equation*}
and consider it to be the measure of the structural complexity of a tree. Informally, the larger the entropy, the more complex is the tree's dendritic structure. The entropy of a tree gives the average number of bits needed to encode it. In order to analyze the entropy's growth rate as $N \rightarrow \infty$, we define the entropy rate $\mathscr{H}_{\infty}$ to be the limit of normalized entropies $H(T_N)/N$, $T_N\in \mathscr{T}_N$ as $N\rightarrow \infty$
\begin{equation*}
\label{entrrate_def}
\mathscr{H}_{\infty} = \lim_{N\rightarrow \infty}\frac{H(T_N)}{N},
\end{equation*}
provided that the limit exists. The entropy rate quantifies per vertex entropy. In other words, for large $N$ the entropy rate gives the average number of bits per vertex required to encode the tree. In fact, for large $N$ there exist an arithmetic coding scheme that encodes a tree with $N$ vertices using about $N\mathscr{H}_{\infty}$ bits \cite{kieffer2009structural}.
Arithmetic coding can get arbitrarily close to the entropy, because it does not convert each vertex separately, but assigns one codeword to the entire tree. The tree can be recreated from this codeword.

The next lemma provides a formula for the entropy of a random planted binary plane tree with $N$ vertices.
\begin{lem}
\label{LemmaTwo}
For a given space $\mathscr{T}_N$, equipped with a uniform distribution, the entropy of a random tree $T_N\in \mathscr{T}_N$ is given by
\begin{eqnarray*}
\label{e13e}
H(T_N) = N-\log_2N-1+\mathcal{O}(\log_2N).
\end{eqnarray*}
\end{lem}
The proof of this result is given in Section \ref{ssec:proofLemmaTwo}.
\begin{cor}
\label{CorOne}
For a given space $\mathscr{T}_N$, equipped with a uniform distribution, the entropy rate is
\begin{eqnarray*}
\label{e14e}
\mathscr{H}_{\infty} = \lim_{N\rightarrow \infty}\frac{H(T_N)}{N} = 1.
\end{eqnarray*}
\end{cor}
The proof of the Corollary \ref{CorOne} is given in Section \ref{ssec:proofCorOne}. Lemma \ref{LemmaTwo} and Corollary \ref{CorOne} demonstrate that for large enough $N$, we need about $N$ bits per tree or about one bit per vertex to encode any tree $T_N \in \mathscr{T}_N$. While presented in a different context, Corollary \ref{CorOne} reaffirms the entropy rate of the maximum entropy model in \cite{kieffer2009structural}.

\subsection{Entropy and entropy rate for space $\mathscr{T}_{K,R}$}

In Section \ref{ssec:HS-ordering} we introduced a space $\mathscr{T}_{N_1\cdots N_K}$ of planted binary plane trees with an arbitrary (but admissible) set Horton-Strahler numbers $N_1,N_2,\cdots,N_K$. 
Quite often, however, observed tree-like structures display geometric decrease of the numbers $N_i$ of elements of Horton-Strahler order $i\geq 1$. This property is known as Horton self-similarity, also referred to as the Horton Law. Formally, the Horton Law states the existence of the limit $\lim_{i\rightarrow\infty}N_i/N_{i+1}=R$, where the quantity $R$ is called the Horton exponent. There are multiple models with broad range of Horton exponents that appear in different scientific areas and have practical importance in a variety of applications \cite{burd2000self,newman1997fractal,peckham1995new,zanardo2013american,milne2017horton}. For example, a perfect binary tree satisfies the Horton law with $R=2$ while the critical binary Galton-Watson tree \cite{burd2000self,peckham1995new,pitman2006combinatorial,shreve1967infinite} satisfies the Horton law with $R=4$. The real river networks have Horton exponent $R$ in a range $(3,5)$ \cite{horton1932drainage,horton1945erosional}, e.g., for Amazon river $R = 4.51$ and for Mississippi river $R = 4.69$. In fact, for many natural tree-like structures $R\in(3,5)$. It was confirmed in hydrology \cite{shreve1966statistical,kirchner1993statistical,peckham1995new,tarboton1996fractal,gupta1998some,rodriguez1997fractal}, biology, and other areas \cite{newman1997fractal}.

Next, we introduce a space of planted binary plane trees that satisfy Horton law with Horton exponent $R$ and examine its entropy rate.

Let $\mathscr{T}_{K,R}$ be a space of planted binary trees with $N$ vertices and the Horton-Strahler numbers $N_k$, $\forall k = \overline{1,K}$ that are defined in a special form as follows
\begin{equation*}
\label{specialHS}
N_k \in \left( R^{K-k} - \alpha^{K-k},R^{K-k} + \alpha^{K-k}\right),
\end{equation*}
where $R, \alpha\in \mathbb{R}$ such that $R>2$ and $\alpha\in(1,R)$. In other words, $N_k \approx R^{K-k}$ with an error $N_k-R^{K-k}$ dominated by the power of an exponent smaller than $R$. It is easy to see that this model satisfies the Horton law with Horton exponent $R$.

\begin{figure}
  \begin{center}
  \includegraphics[scale=0.6]{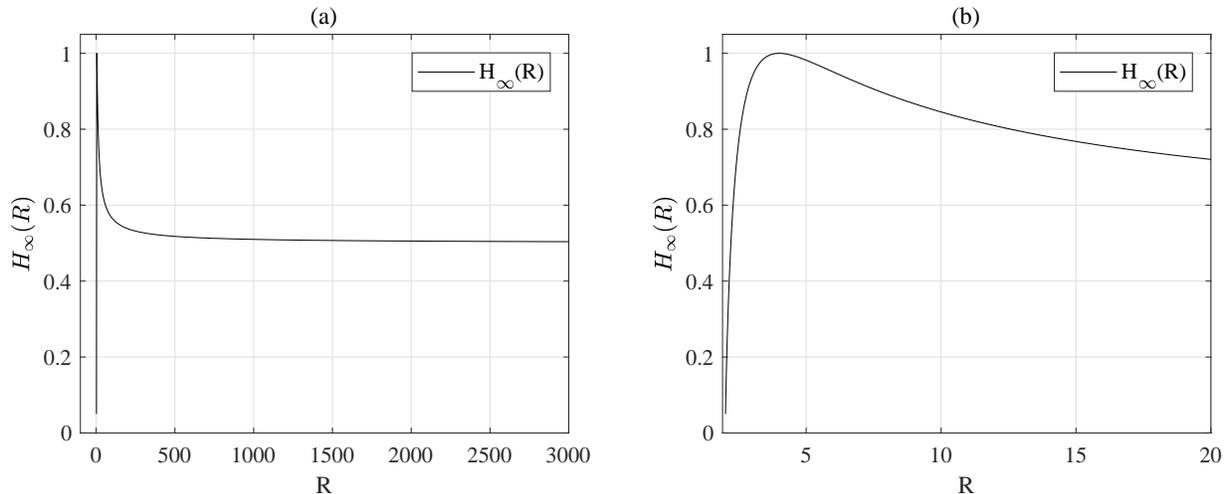}\\
  \end{center}
  \caption{Entropy rate $\mathscr{H}_{\infty}(R)$ for (a) $R\in (0,3000]$ and (b) $R\in (0,20]$. The maximum of $\mathscr{H}_{\infty}(R)$ is attained at $R=4$.}
  \label{plot-entropy-rate}
\end{figure}

\begin{thm}
\label{entrRateSpCase}
For a given space $\mathscr{T}_{K,R}$, equipped with a uniform distribution, the entropy rate is given by
\begin{equation*}
\label{enrtRateModel}
\mathscr{H}_{\infty}(R) 1-\frac{1-H\left(2/R\right)}{2-2/R},
\end{equation*}
where $H(z) = -z\log_2z -(1-z)\log_2(1-z)$ is a binary entropy of $z$.
\end{thm}
The proof of Theorem \ref{entrRateSpCase} is given in Section \ref{ssec:proofTHentropyrateR}. In Figure \ref{plot-entropy-rate} we depict the entropy rate $\mathscr{H}_{\infty}(R)$ for a range of values of $R$. The entropy rate is zero for $R=2$ because the dendritic structure of a perfect planted binary plane tree is predetermined for any $N$. Note that, when $R$ is allowed to grow, the entropy rate converges to $1/2$. More precisely,
\begin{equation*}
\label{MainFormulaLimitR}
\lim_{R\rightarrow\infty}\mathscr{H}_{\infty}(R) = \lim_{R\rightarrow\infty}\left(1-\frac{1-H\left(2/R\right)}{2-2/R}\right) = \frac{1}{2}.
\end{equation*}
Thus, for large $R$ and $N$ one would need about $N/2$ bits to decode the entire tree. It would be interesting to explain why trees with large enough $R$ require less bits to encode them than the trees with $R=4$.

For $R=4$ the entropy rate attains its maximal value $1$. 
Recall that the critical binary Galton-Watson model has parameter $R=4$. It was noticed that the Horton exponents for real rivers are different from the theoretical parameter $R=4$ \cite{peckham1995new}. For example, for Amazon river $R = 4.51$. Consequently, entropy rate for Amazon river is $0.9941$. A natural question to ask would be: What physical phenomenon does cause the nonoptimality of entropy rate of the rivers? A possible explanation of this phenomenon is given in \cite{tejedor2017entropy}: although river deltas adjust their configurations to maximize the entropy, this maximization happens within local feasibility constraints, thus global maximum is not achieved.

\section{Conclusion}
\label{sec:conclusion}

This work was motivated by the growing interest in statistical and complexity characteristics of tree-like structures. We considered several spaces of planted binary plane trees and examined structural complexity of those trees. Specifically, we calculated the number of planted binary plane trees with particular Horton-Strahler orders. We defined and evaluated the entropy for a space of planted binary plane trees with $N$ vertices. We introduced the entropy rate measure in order to explain the long term behavior of growing tree model and find closed-form formulas for the entropy rate for a space of planted binary plane trees with $N$ vertices. Moreover, we found entropy rate for a space of planted binary plane trees that satisfy the Horton Law with Horton exponent $R$. The author is currently working on extending these results under different forms of trees self-similarity.

\section{Appendix}
\label{sec:Appendix}

\subsection{Auxiliary Lemma}
\label{ssec:Lemma0}

\begin{lem}
\label{lemma_0}
For any $n$ and $k$ such that $n\geq k\geq 1$ the following asymptotic approximation is true
\begin{equation}
\label{23}
\log_2{n \choose k} = nH\left(\frac{k}{n}\right)+\mathcal{O}(\log_2n),
\end{equation}
where ${n \choose k} = \frac{n!}{k!(n-k)!}$ and $H(z) = -z\log_2z -(1-z)\log_2(1-z)$ is a binary entropy of $z$.
\end{lem}

\proof
Using the Stirling's approximation $\log_2n! = n\log_2n-(\log_2e)n+\mathcal{O}(\log_2n)$ we obtain the required approximation as follows
\begin{eqnarray}
\label{26}
\log_2 {n \choose k} &=& \log_2\left(\frac{n!}{k!(n-k)!}\right)\\
&=& n\log_2n-(\log_2e)n-k\log_2k+(\log_2e)k-(n-k)\log_2(n-k)\\
&+&(\log_2e)(n-k) + \mathcal{O}(\log_2n)-\mathcal{O}(\log_2k)-\mathcal{O}(\log_2(n-k))\\
&=& n\log_2n-k\log_2k-(n-k)\log_2(n-k)+k\log_2n-k\log_2n+\mathcal{O}(\log_2n)\\
&=& n\left(-\frac{k}{n}\log_2\left(\frac{k}{n}\right)-\left(1-\frac{k}{n}\right)\log_2\left(1-\frac{k}{n}\right)\right)+\mathcal{O}(\log_2n)\\
&=& nH\left(\frac{k}{n}\right)+\mathcal{O}(\log_2n).
\end{eqnarray}
\endproof

\subsection{Proof of Lemma \ref{CountLemmaOne}}
\label{ssec:proofCountLemmaOne}

\proof
We prove this theorem by providing a method to construct and count trees with fixed Horton-Strahler numbers. We start by introducing a few helpful definitions.


For a given tree we define the $\mathit{main}$ $\mathit{frame}$ (also know as a skeleton in related publications) to be the minimal subtree of the same order with the same root. Each branch of order $i+1$ is obtained by merging two $\mathit{necessary}$ $\mathit{frames}$ of order $i$, $\forall i = \overline{1,K-1}$. All other frames are $\mathit{extra}$ $\mathit{frames}$. Thus, given a set of Horton-Strahler numbers $N_i$ such that $N_i\geq 2N_{i+1}$, $\forall i = \overline{1,K-1}$, the number of necessary frames of order $i$ is $L_i = 2N_{i+1}$ and the number of extra frames of order $i$ is $M_i = N_i-L_i$.


\begin{figure}
  \begin{center}
  \includegraphics[scale=0.4]{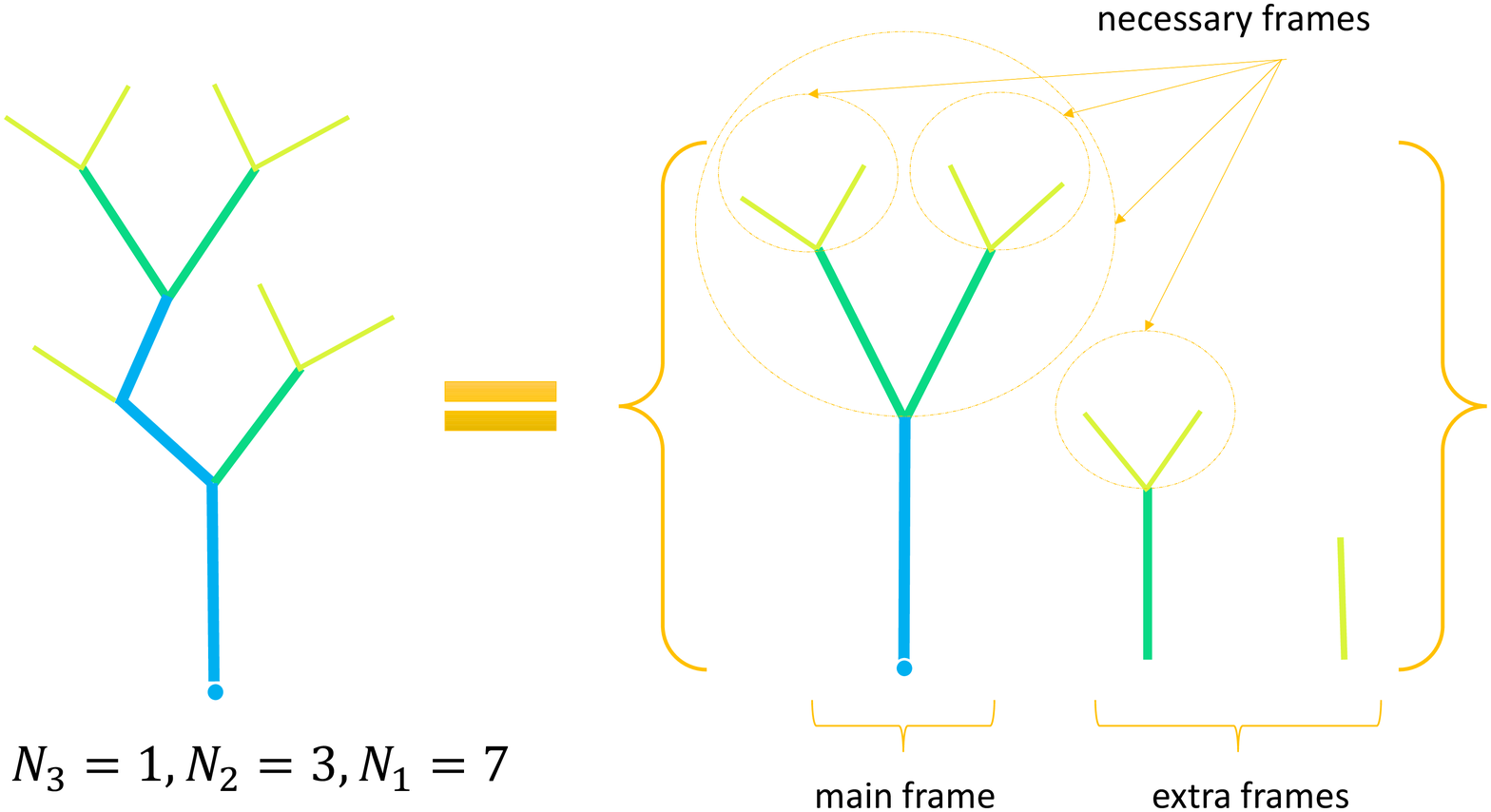}\\
  \end{center}
  \caption{An example of a planted binary plane tree of order $K = 3$ with $N_3 = 1, N_2 = 3,$ $N_1 = 7$. There are one main frame of order 3, two necessary frames of order 2 ($L_2 = 2$), six necessary frames of order 1 ($L_1 = 6$), one extra frame of order 2 ($M_2 = 1$), and one extra frame of order 1 ($M_1 = 1$). This tree is constructed by attaching extra frames of orders $2$ and $1$ to the main frame of order $3$. Branches of order $3$ are depicted in blue, branches of order $2$ in green, and branches of order $1$ in pear.}
  \label{neframes}
\end{figure}



\begin{figure}
  \begin{center}
  \includegraphics[scale=0.4]{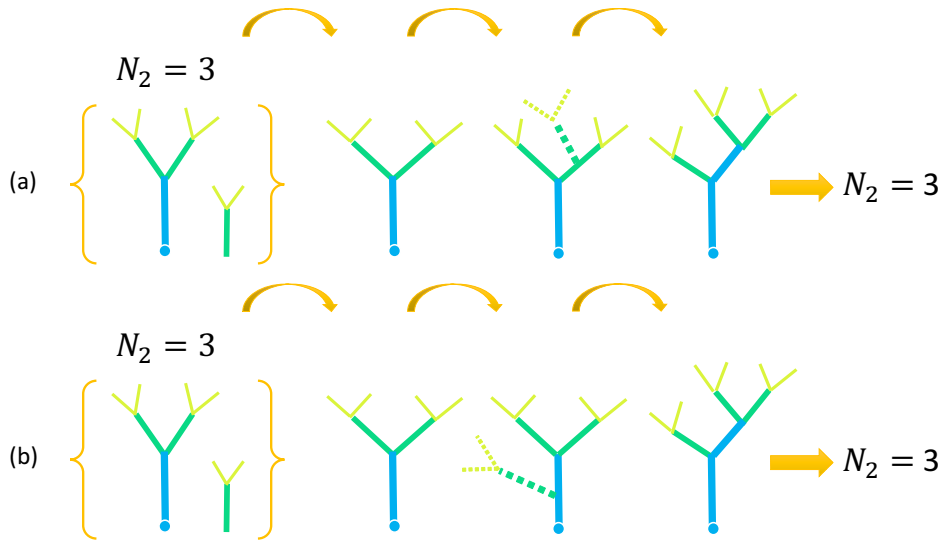}\\
  \end{center}
  \caption{An example of attaching an extra frame to the branch of (a) the same and (b) a higher order. In (a) we attach an extra frame of order $2$ to the branch of order $2$. In (b) we attach an extra frame of order $2$ to the branch of order $3$. The resulting trees are identical and have correct Horton-Strahler numbers. Branches of order $3$ are depicted in blue, branches of order $2$ in green, and branches of order $1$ in pear.}
  \label{h_order}
\end{figure}

To illustrate the notion of necessary and extra frames, consider a planted binary plane tree of order $K = 3$ depicted in Figure \ref{neframes}. Note that each extra frame of order $i$ is attached to the branch of higher order $j>i$ because all other ways to attach extra frames will result in either non-binary trees or in binary trees with incorrect Horton-Strahler numbers and incorrect number of vertices. For example, no extra frame can be attached to the root node of the main frame, since the root node should be of degree $1$. Also extra frame can not be attached to the leaf vertices or to the internal vertices of the main frame; otherwise it will result in a non-binary tree. Moreover, attaching an extra frame of order $i$ to the branch of lower order $j<i$ will result in a tree with incorrect number of branches of order $i$: instead of $N_i$ the tree will have $N_i-1$ branches of order $i$. Finally, attaching an extra frame of order $i$ to the branch of the same order $j=i$ results in a tree that is redundant to the tree constructed by attaching an extra frame of order $i$ to the branch of higher order. See Figure \ref{h_order} for an example. Therefore, to find the total number of planted binary plane trees of order $K$ with $N=2N_1$ vertices and given Horton-Strahler numbers $N_1, N_2, \cdots, N_K$, we should start with a main frame of order $K$ and then count all possible ways we can attach all extra frames to the branches of higher orders, starting with the extra frames of order $K-1$, followed by the extra frames of order $K-2$, and so on. Extra frames of order $1$ will be attached at the end.

We start with the main frame of order $K$. Denote $T_{K-1\rightarrow K}$ to be the number of trees we obtain by attaching $M_{K-1}$ extra frames of order $K-1$ to one branch of order $K$ of the main frame. In general, the number of ways to place $n$ identical objects into $k$ different positions is given by the formula
\begin{equation}
\label{n-ident-obj-onto-k-diff-boxes}
{n+k-1 \choose k-1} = \frac{(n+k-1)!}{(k-1)!n!}.
\end{equation}
We also need to take into account that each extra frame can be attached to the middle point of a branch either form the left or from the right. Therefore, $T_{K-1\rightarrow K}$ can be calculated as follows
\begin{equation}
\label{e1}
T_{K-1\rightarrow K} = 2^{M_{K-1}}{N_K+M_{K-1}-1 \choose N_K-1} = 2^{N_{K-1}-2N_K}{N_{K-1}-2 \choose N_K-1}.
\end{equation}

Note that when we attach $M_{K-1}$ extra frames of order $K-1$ to one branch of order $K$, we brake the branch of order $K$ into $N_K+M_{K-1}$ edges. The Horton-Strahler number does not change: there is still one branch of order $K$, i.e., $N_K = 1$, but it consists of $N_K+M_{K-1}$ edges of order $K$. For example in  Figure \ref{h_order} (b), attachment of an extra frame of order $2$ to the branch of order $3$, broke the branch of order $3$ into two edges.

Next, denote $T_{K-2\rightarrow K-1,K}$ to be the the number of different trees we obtain by attaching extra frames of order $K-2$ to branches of higher orders $K-1$ and $K$. There are $N_K+M_{K-1}$ edges of order $K$ and $N_{K-1}$ branches of order $K-1$. Thus, there are $k = N_K+M_{K-1}+N_{K-1} = 2N_{K-1} - 1$ edges of orders $K$ and $K-1$ to which we can attach extra frames of order $K-2$. By using formula (\ref{n-ident-obj-onto-k-diff-boxes}) and considering that each extra frame can be attached either from the left or from the right we obtain $T_{K-2\rightarrow K-1,K}$ as follows
\begin{eqnarray}
\label{e2}
T_{K-2\rightarrow K-1,K} &=& 2^{N_{K-2}-2N_{K-1}}{N_{K-2}-2 \choose 2N_{K-1}-2}.
\end{eqnarray}

Consider now an intermediate step. Let $T_{i\rightarrow i+1,\cdots,K}$ be the number of trees that we obtain by attaching $M_i$ extra frames of order $i$ to the branches of orders $i+1, i+2, \cdots,K$. Note that there are now
\begin{eqnarray}
\label{intermMi} \nonumber
k &=& N_K+M_{K-1}+N_{K-1}+M_{K-2}+N_{K-2}+\cdots +M_{i+1}+N_{i+1}\\ \nonumber
&=& N_K+N_{K-1}-2N_{K}+N_{K-1}+N_{K-2}-2N_{K-1}+N_{K-2}+\cdots +N_{i+1}-2N_{i+2}+N_{i+1}\\
&=& 2N_{i+1}-N_K = 2N_{i+1}-1
\end{eqnarray}
edges of orders $i+1, i+2, \cdots, K$, to which we can attach extra frames of order $i$. Thus, using formula (\ref{n-ident-obj-onto-k-diff-boxes}) and considering that each extra frame can be attached either from the left or from the right we obtain $T_{i\rightarrow i+1,\cdots,K}$ as follows
\begin{equation}
\label{e3}
T_{i\rightarrow i+1,\cdots,K} = 2^{N_{i}-2N_{i+1}}{N_{i}-2 \choose 2N_{i+1}-2}.
\end{equation}
Equation (\ref{e3}) provides a general formula for the terms $T_{i\rightarrow i+1,\cdots, K}$, $\forall i = \overline{K-2,1}$.

Note now that for every possible attachment of extra frames of order $i+1$ there are $T_{i\rightarrow i+1,\cdots,K}$ possible attachments of extra frames of order $i,$ $\forall i = \overline{K-1,1}$. Thus, using the multiplication principle of combinatorics, we obtain the total number of planted binary plane trees of order $K$ with particular Horton-Strahler numbers $N_1, N_2, \cdots, N_K$ and $N = 2N_1$ vertices as follows
\begin{eqnarray}
\label{e6} \nonumber
|\mathscr{T}_{N_1\cdots N_K}| &=& \prod_{i=K-1}^{1}T_{i\rightarrow i+1,\cdots,K} =\prod_{i=K-1}^{1}2^{N_{i}-2N_{i+1}}{N_{i}-2\choose 2N_{i+1}-2} \\ \nonumber
&=& 2^{\sum_{i=1}^{K-1}\left(N_{i}-2N_{i+1}\right)}\prod_{i=1}^{K-1}{N_{i}-2 \choose 2N_{i+1}-2} =  2^{N_1-1-\sum_{i=1}^{K-1}N_{i+1}}\prod_{i=1}^{K-1}{N_{i}-2 \choose 2N_{i+1}-2}.
\end{eqnarray}
\endproof

\subsection{Proof of Lemma \ref{LemmaTwo}}
\label{ssec:proofLemmaTwo}

\proof
First note that the probability of a random tree $T_N\in \mathscr{T}_N$ is $P(T_N) = 1/\mathcal{C}_{n-1}$ since we assume the uniform distribution of trees in $\mathscr{T}_N$. Thus, the entropy is given as follows
\begin{equation}
\label{entropy_1}
H(T_N) = -\mathbb{E}[\log_2P(T_N)] = -\sum_{i = 1}^{\mathcal{C}_{n-1}}\frac{1}{\mathcal{C}_{n-1}}\log_2\frac{1}{\mathcal{C}_{n-1}} = \log_2\mathcal{C}_{n-1}.
\end{equation}
Note that term $\log_2 \mathcal{C}_{n-1}$ can be rewritten in the following way
\begin{equation}
\label{e12}
\mathcal{C}_{n-1} = \frac{(2n-2)!}{n!(n-1)!}= \frac{\left(2\left(\frac{N}{2}-1\right)\right)!}{\left(\frac{N}{2}\right)!\left(\frac{N}{2}-1\right)!}=
\frac{\left(2\left(\frac{N}{2}-1\right)\right)!}{\left(\frac{N}{2}\right)\left(\left(\frac{N}{2}-1\right)!\right)^2} = \frac{2}{N}{N-2\choose \frac{N}{2}-1},
\end{equation}
where we used the fact that $n = \frac{N}{2}$.
Using the results of Lemma \ref{lemma_0}, given in Section \ref{ssec:Lemma0}, we obtain the entropy of $T_N$ as follows
\begin{eqnarray}
\label{e13} \nonumber
H(T_N) &=& \log_2\left[\frac{2}{N}{N-2\choose \frac{N}{2}-1}\right]= 1-\log_2N+2\left(\frac{N}{2}-1\right)H\left(\frac{1}{2}\right)+\mathcal{O}(\log_2N)\\
&=& 1-\log_2N+N-2 +\mathcal{O}(\log_2N) = N-\log_2N-1+\mathcal{O}(\log_2N).
\end{eqnarray}
\endproof

\subsection{Proof of Corollary \ref{CorOne}}
\label{ssec:proofCorOne}

\proof
Dividing $H(T_N)$ by $N$ and taking the limit as $N \rightarrow \infty$ we obtain the entropy rate as follows
\begin{eqnarray}
\label{e14}
\mathscr{H}_{\infty}(\mathscr{T}_N) = \lim_{N\rightarrow \infty}\frac{H(T_N)}{N} = \lim_{N\rightarrow \infty}\frac{N-\log_2N-1+\mathcal{O}(\log_2N)}{N} = 1.
\end{eqnarray}
\endproof

\subsection{Proof of Theorem \ref{entrRateSpCase}}
\label{ssec:proofTHentropyrateR}

We begin with two auxiliary lemmas that will be used in the proof of Theorem \ref{entrRateSpCase}.

\begin{lem}
\label{lemma_hlp}
Given parameters $K$ and $R$, the following properties hold in space $\mathscr{T}_{K,R}$
\begin{enumerate}
  \item $N_k\geq 2^{K-k}$, $\forall k = \overline{1,K}$,
  \item $N_K=1$,
  \item $\lim_{K \rightarrow \infty}\frac{N_k}{N_1} = R^{1-k}$,
  \item $\lim_{K\rightarrow \infty}\frac{N_1}{R^{K-1}} = 1,$
  \item $\lim_{K\rightarrow \infty}\frac{N}{R^{K-1}} = 2$.
\end{enumerate}
\end{lem}

\proof
\begin{enumerate}
  \item Using the fact that $\forall k = \overline{1,K-1}$ the Horton-Strahler numbers satisfy $N_k\geq 2N_{k+1}$ and $R>2$, we conclude that $N_k\geq 2N_{k+1} = 2R^{K-(k+1)}>2^{K-k}$.
  \item Let $k = K$ then $N_K-1\in (-1,1)$. Since $N_K\in Z$, then $N_K=1$.
  \item $\lim_{K \rightarrow \infty}\frac{N_k}{N_1} = \lim_{K \rightarrow \infty}\frac{R^{K-k}\pm\alpha^{K-k}}{R^{K-1}\pm\alpha^{K-1}} = \lim_{K \rightarrow \infty}\frac{R^{1-k}\pm\alpha^{1-k}\left(\frac{\alpha}{R}\right)^{K-1}}{1\pm\left(\frac{\alpha}{R}\right)^{K-1}}=R^{1-k}$.
  \item $\lim_{K\rightarrow \infty}\frac{N_1}{R^{K-1}} = \lim_{K\rightarrow \infty}\frac{R^{K-1}\pm \alpha^{K-1}}{R^{K-1}} = \lim_{K \rightarrow \infty}\left(1\pm\frac{\alpha^{K-1}}{R^{K-1}}\right) =1$.
  \item Since for any planted binary plane tree $N=2N_1$, then $\lim_{K\rightarrow \infty}\frac{N}{R^{K-1}} = \lim_{K\rightarrow \infty}\frac{2N_1}{R^{K-1}} = 2$.
\end{enumerate}
\endproof

\begin{lem}
\label{lemma_hlp2}
Let $\mathscr{T}_{N_1,N_2,\cdots,N_K,R}\subset\mathscr{T}_{K,R}$ be a space of planted binary trees with particular set of Horton-Strahler numbers $N_1,N_2,\cdots,N_K$, such that $N_k=R^{K-k}\pm \alpha^{K-k}$, $k=\overline{1,K}$. Then
\begin{equation}\nonumber
\lim_{N \rightarrow \infty}\frac{\log_2|\mathscr{T}_{N_1,N_2,\cdots,N_K,R}|}{N}= 1-\frac{1-H\left(2/R\right)}{2-2/R}.
\end{equation}
\end{lem}
\proof
We begin the proof by using the results of Lemma \ref{CountLemmaOne} that gives us the number of planted binary trees with given Horton-Strahler numbers $N_1, N_2,\cdots,N_K$. Thus,
\begin{eqnarray}
\label{twolim}\nonumber
\lim_{N \rightarrow \infty}\frac{\log_2|\mathscr{T}_{N_1,N_2,\cdots,N_K,R}|}{N}&=& \lim_{N \rightarrow \infty}\frac{1}{N}\log_2\left(2^{N_1-1-\sum_{i=1}^{K-1}N_{i+1}}\prod_{i = 1}^{K-1} {N_i-2 \choose 2N_{i+1}-2}\right)\\\nonumber
&=& \lim_{N \rightarrow \infty}\frac{1}{N}\left[N_1-1-\sum_{i=1}^{K-1}N_{i+1}\right]+ \lim_{N \rightarrow \infty}\frac{1}{N}\left[\sum_{i=1}^{K-1}\log_2{N_i-2 \choose 2N_{i+1}-2}\right].
\end{eqnarray}
Note that the term ${N_i-2 \choose 2N_{i+1}-2}$ can be
rewritten in the following way
\begin{equation}\nonumber
\label{e18}
{N_i-2 \choose 2N_{i+1}-2} =  \frac{\left(N_i-2\right)!}{\left(2N_{i+1}-2\right)!\left(N_i-2N_{i+1}\right)!} = {N_i\choose 2N_{i+1}}\frac{\left(2N_{i+1}-2\right)\left(2N_{i+1}-1\right)}{\left(N_i-2\right)\left(N_i-1\right)}.
\end{equation}
Therefore, $\lim_{N \rightarrow \infty}\frac{\log_2|\mathscr{T}_{N_1,N_2,\cdots,N_K,R}|}{N} = $
\begin{eqnarray}
\label{twolim2} \nonumber
&=& \lim_{N \rightarrow \infty}\frac{1}{N}\left[N_1-1-\sum_{i=1}^{K-1}N_{i+1}\right]+ \lim_{N \rightarrow \infty}\frac{1}{N}\sum_{i = 1}^{K-1}\log_2{N_i\choose 2N_{i+1}}\\
&+&\lim_{N \rightarrow \infty}\frac{1}{N}\sum_{i = 1}^{K-1}\log_2\frac{2N_{i+1}-2}{N_i-2} + \lim_{N \rightarrow \infty}\frac{1}{N}\sum_{i =1}^{K-1}\log_2\frac{2N_{i+1}-1}{N_i-1}.
\end{eqnarray}
We consider each of the four limits in (\ref{twolim}) separately. Starting with the first limit, we notice that term $N_1-1-\sum_{i=1}^{K-1}N_{i+1}$ can be rewritten in the following way
\begin{eqnarray}
\label{e17}\nonumber
N_1-1-\sum_{i=1}^{K-1}N_{i+1} &=& 2N_1-1-\sum_{i=1}^{K}N_i = N-1-\sum_{i=1}^{K}\left(R^{K-i}\pm \alpha^{K-i}\right)\\\nonumber
&=& N-1-R^{K}\sum_{i=1}^{K}R^{-i}-(\pm1)\alpha^{K}\sum_{i=1}^{K}\alpha^{-i}\\
&=& N-1-\frac{R^K-1}{R-1}-(\pm 1)\frac{\alpha^{K}-1}{\alpha-1}.
\end{eqnarray}
Thus, dividing equation (\ref{e17}) by $N$ and taking the limit as $N\rightarrow \infty$ we find the value of the first limit in (\ref{twolim})
\begin{eqnarray}
\label{e41}\nonumber
\lim_{N \rightarrow \infty}\frac{1}{N}\left[N_1-1-\sum_{i=1}^{K-1}N_{i+1}\right]&=& \lim_{N \rightarrow \infty}\frac{1}{N} \left[N-1-\frac{R^K-1}{R-1}-(\pm 1)\frac{\alpha^{K}-1}{\alpha-1}\right]\\
& = & 1 -\lim_{N \rightarrow \infty}\frac{1}{N}\frac{R^K-1}{R-1}= 1-\frac{R/2}{R-1},
\end{eqnarray}
where the last equation is obtained using result 5) of Lemma \ref{lemma_hlp}.\\
Consider now the second limit in equation (\ref{twolim}). Using the result of Lemma \ref{lemma_0}, provided in Section \ref{ssec:Lemma0}, we can rewrite the second term as follows
\begin{eqnarray}
\label{e27}\nonumber
\lim_{N \rightarrow \infty}\frac{1}{N}\sum_{i = 1}^{K-1}\log_2{N_i \choose 2N_{i+1}}&=& \lim_{N \rightarrow \infty}\frac{1}{N}\sum_{i = 1}^{K-1}\left[N_iH\left(\frac{2N_{i+1}}{N_i}\right)+\mathcal{O}(\log_2N_i)\right].
\end{eqnarray}
To examine the term $\sum_{i = 1}^{K-1}N_iH\left(\frac{2N_{i+1}}{N_i}\right)$, we break it into two sums
\begin{eqnarray}
\label{e33}
\sum_{i = 1}^{K-1}N_iH\left(\frac{2N_{i+1}}{N_i}\right)&=& \sum_{i = 1}^{K^{'}-1}N_iH\left(\frac{2N_{i+1}}{N_i}\right)+\sum_{i =K^{'}}^{K-1}N_iH\left(\frac{2N_{i+1}}{N_i}\right),
\end{eqnarray}
where $K^{'} = \left\lceil\frac{K}{2}\right\rceil$.\\
Consider the first sum in equation (\ref{e33}). Using the fact that $1\leq i\leq K^{'}-1$, we obtain the following upper bound on term $\frac{2N_{i+1}}{N_i}$
\begin{equation*}
\frac{2N_{i+1}}{N_i} \leq  2\frac{R^{K-(i+1)} +\alpha^{K-(i+1)}}{R^{K-i}-\alpha^{K-i}}= \frac{2}{R}\times\frac{1+\left(\frac{\alpha}{R}\right)^{K-(i+1)}}{1-\left(\frac{\alpha}{R}\right)^{K-i}}
\leq \frac{2}{R}\times\frac{1+\left(\frac{\alpha}{R}\right)^{K-K^{'}}}{1-\left(\frac{\alpha}{R}\right)^{K-1}}.
\end{equation*}
Thus,
\begin{equation}
\label{ggr1}
\frac{2N_{i+1}}{N_i} \leq \frac{2}{R}\left(1+\mathcal{O}\left(\left(\frac{\alpha}{R}\right)^{\frac{K}{2}}\right)\right). \end{equation}
In a similar fashion we obtain a lower bound on term $\frac{2N_{i+1}}{N_i}$
\begin{equation}
\label{ggr2}
\frac{2}{R}\left(1-\mathcal{O}\left(\left(\frac{\alpha}{R}\right)^{\frac{K}{2}}\right)\right)
\leq \frac{2N_{i+1}}{N_i}.
\end{equation}
Combining two bounds in (\ref{ggr1}) and (\ref{ggr2}) together, we obtain
\begin{equation*}
\frac{2}{R}\left(1-\mathcal{O}\left(\left(\frac{\alpha}{R}\right)^{\frac{K}{2}}\right)\right)
\leq \frac{2N_{i+1}}{N_i}\leq \frac{2}{R}\left(1+\mathcal{O}\left(\left(\frac{\alpha}{R}\right)^{\frac{K}{2}}\right)\right).
\end{equation*}
Since the entropy function $H(\cdot)$ has bounded derivative in any small enough closed neighborhood around $\frac{2}{R}$, we can bound term $H\left(\frac{2N_{i+1}}{N_{i}}\right)$ as follows
\begin{equation}
\label{e35}
H\left(\frac{2}{R}\right)\left(1-\mathcal{O}\left(\left(\frac{\alpha}{R}\right)^{\frac{K}{2}}\right)\right)
\leq H\left(\frac{2N_{i+1}}{N_i}\right)\leq H\left(\frac{2}{R}\right)\left(1+\mathcal{O}\left(\left(\frac{\alpha}{R}\right)^{\frac{K}{2}}\right)\right).
\end{equation}
Using similar arguments, we obtain the following bounds on the term $N_i$
\begin{equation}
\label{e34}
R^{K-i}\left(1-\left(\frac{\alpha}{R}\right)^{\frac{K}{2}}\right)\leq N_i\leq R^{K-i}\left(1+\left(\frac{\alpha}{R}\right)^{\frac{K}{2}}\right).
\end{equation}
Thus, combining formulas (\ref{e35}) and (\ref{e34}) we get
\begin{equation*}
R^{K-i}H\left(\frac{2}{R}\right)\left(1-\mathcal{O}\left(\left(\frac{\alpha}{R}\right)^{\frac{K}{2}}\right)\right)
\leq N_iH\left(\frac{2N_{i+1}}{N_i}\right)\leq R^{K-i}H\left(\frac{2}{R}\right)\left(1+\mathcal{O}\left(\left(\frac{\alpha}{R}\right)^{\frac{K}{2}}\right)\right).
\end{equation*}
Therefore,
\begin{eqnarray}\nonumber
\label{y1}
\sum_{i=1}^{K^{'}-1}N_iH\left(\frac{2N_{i+1}}{N_i}\right)&\leq& \sum_{i=1}^{K^{'}-1}R^{K-i}H\left(\frac{2}{R}\right)\left(1+\mathcal{O}\left(\left(\frac{\alpha}{R}\right)^{\frac{K}{2}}\right)\right)\\\nonumber
&\leq& R^{K}H\left(\frac{2}{R}\right)\left(\sum_{i=1}^{K^{'}-1}R^{-i}\right)\left(1+\mathcal{O}\left(\left(\frac{\alpha}{R}\right)^{\frac{K}{2}}\right)\right)\\
&=& R^{K-1}H\left(\frac{2}{R}\right)\frac{1-1/R^{K^{'}-1}}{1-1/R}\left(1+\mathcal{O}\left(\left(\frac{\alpha}{R}\right)^{\frac{K}{2}}\right)\right).
\end{eqnarray}
Taking the limit as $N\rightarrow \infty$ in (\ref{y1}), we obtain
\begin{eqnarray}\nonumber
\label{t1}
\lim_{N\rightarrow \infty}\frac{1}{N}\sum_{i=1}^{K^{'}-1}N_iH\left(\frac{2N_{i+1}}{N_i}\right)&\leq& \lim_{N\rightarrow \infty}\frac{R^{K-1}}{N}H\left(\frac{2}{R}\right)\frac{1-1/R^{K^{'}-1}}{1-1/R}\left(1+\mathcal{O}\left(\left(\frac{\alpha}{R}\right)^{\frac{K}{2}}\right)\right)\\
&=& H\left(\frac{2}{R}\right)\frac{R/2}{R-1}.
\end{eqnarray}
Similarly, we show that
\begin{eqnarray*}\nonumber
\sum_{i=1}^{K^{'}-1}N_iH\left(\frac{2N_{i+1}}{N_i}\right)&\geq& R^{K-1}H\left(\frac{2}{R}\right)\frac{1-1/R^{K^{'}-1}}{1-1/R}\left(1-\mathcal{O}\left(\left(\frac{\alpha}{R}\right)^{\frac{K}{2}}\right)\right)\\
\end{eqnarray*}
and hence
\begin{eqnarray}\nonumber
\label{t2}
\lim_{N\rightarrow \infty}\frac{1}{N}\sum_{i=1}^{K^{'}-1}N_iH\left(\frac{2N_{i+1}}{N_i}\right)&\geq& \lim_{N\rightarrow \infty}\frac{R^{K-1}}{N}H\left(\frac{2}{R}\right)\frac{1-1/R^{K^{'}-1}}{1-1/R}\left(1-\mathcal{O}\left(\left(\frac{\alpha}{R}\right)^{\frac{K}{2}}\right)\right)\\
&=& H\left(\frac{2}{R}\right)\frac{R/2}{R-1}.
\end{eqnarray}
Thus, combining formulas (\ref{t1}) and (\ref{t2}), we obtain
\begin{equation}
\label{eqw1}
\lim_{N\rightarrow \infty}\frac{1}{N}\sum_{i=1}^{K^{'}-1}N_iH\left(\frac{2N_{i+1}}{N_i}\right)= H\left(\frac{2}{R}\right)\frac{R/2}{R-1}.
\end{equation}
Consider now the second term in equation (\ref{e33}), where $K^{'}\leq i\leq K-1$. Using the fact that the entropy function is always bounded by 1 from above, we obtain
\begin{eqnarray}\nonumber
\label{t3}
\sum_{i=K^{'}}^{K-1}N_iH\left(\frac{2N_{i+1}}{N_i}\right)&\leq& \sum_{i=K^{'}}^{K-1}N_i \leq  \sum_{i=K^{'}}^{K-1}\left(R^{K-i}+\alpha^{K-i}\right) = \sum_{k=1}^{K-K^{'}}\left(R^{k}+\alpha^{k}\right)\\
&=& R\frac{R^{K-K^{'}}-1}{R-1}+\alpha\frac{\alpha^{K-K^{'}}-1}{\alpha-1} \leq \frac{R^{K-K^{'}+1}}{R-1}+\frac{\alpha^{K-K^{'}+1}}{\alpha-1}.
\end{eqnarray}
Hence, dividing formula (\ref{t3}) by $N$ and taking a limit as $N\rightarrow\infty$, we obtain
\begin{eqnarray*}\nonumber
\lim_{N\rightarrow\infty}\frac{1}{N}\sum_{i=K^{'}}^{K-1}N_iH\left(\frac{2N_{i+1}}{N_i}\right) &=& \lim_{N\rightarrow\infty}\frac{1}{N}\left(\frac{R^{K-K^{'}+1}}{R-1}+\frac{\alpha^{K-K^{'}+1}}{\alpha-1}\right)\\
&=&\lim_{N\rightarrow\infty}\left(\frac{R^{K-1}}{N}\left(\frac{R^{2-K^{'}}}{R-1}\right)+\frac{\alpha^{K-1}}{N}\left(\frac{\alpha^{2-K^{'}}}{R-1}\right)\right)=0.
\end{eqnarray*}
Therefore, the second limit in equation (\ref{twolim}) is
\begin{equation}
\label{e40}
\lim_{N \rightarrow \infty}\frac{1}{N}\sum_{i = 1}^{K-1}\log_2{N_i\choose 2N_{i+1}}=H\left(\frac{2}{R}\right)\frac{R/2}{R-1}.
\end{equation}
Consider now the third term in equation (\ref{twolim}). Note that $\frac{2N_{i+1}-2}{N_i-2}\leq 1$, since $N_i\geq 2N_{i+1}$.  
Moreover, $\forall i = \overline{1,K-2}$ $N_{i+1}\geq N_{K-1}\geq 2N_K=2$ and $N_i-2\leq N_i\leq N_{1} \leq R^{K-1}+\alpha^{K-1}$.
Therefore,
\begin{equation*}
\label{e39}
1 \geq \frac{2N_{i+1}-2}{N_i-2} \geq \frac{2\times 2-2}{R^{K-1}+\alpha^{K-1}-2} \geq \frac{2}{R^{K-1}+\alpha^{K-1}}\geq \frac{2}{2R^{K-1}} = \frac{1}{R^{K-1}}.
\end{equation*}
Thus,
\begin{equation*}
0\geq \log_2\left(\frac{2N_{i+1}-2}{N_i-2}\right) \geq (K-1)\log_2\left(\frac{1}{R}\right)
\end{equation*}
and
\begin{equation}
\label{t4}
0\geq \sum_{i=1}^{K-1}\log_2\left(\frac{2N_{i+1}-2}{N_i-2}\right) \geq (K-1)^2\log_2\left(\frac{1}{R}\right).
\end{equation}
Thus, dividing all sides of inequality in (\ref{t4}) by $N$ and taking a limit as $N \rightarrow \infty$, we show the third term in equation (\ref{twolim}) is equal to zero
\begin{equation}
\label{e42}
0\geq \lim_{N\rightarrow \infty}\frac{1}{N}\sum_{i=1}^{K-1}\log_2\left(\frac{2N_{i+1}-2}{N_i-2}\right) \geq \lim_{N\rightarrow \infty}\frac{(K-1)^2}{N}\log_2\left(\frac{1}{R}\right) = 0,
\end{equation}
where the equation on the right hand side follows from Lemma \ref{lemma_hlp}.
Similarly, we show that the fourth term in equation (\ref{twolim}) is equal to zero
\begin{equation}
\label{e30}
\lim_{N\rightarrow \infty}\frac{1}{N}\sum_{i = 1}^{K-1}\log_2\left(\frac{2N_{i+1}-1}{N_i-1}\right) = 0.
\end{equation}
Thus, combining formulas (\ref{e41}), (\ref{e40}), (\ref{e42}), and (\ref{e30}) we find
\begin{eqnarray*}
\lim_{N \rightarrow \infty}\frac{\log_2|\mathscr{T}_{N_1,N_2,\cdots,N_K,R}|}{N}&=& H\left(\frac{2}{R}\right)\frac{R/2 }{R-1}+\left(1-\frac{R/2}{R-1}\right)= 1-\frac{1-H\left(2/R\right)}{2-2/R}.
\end{eqnarray*}
\endproof

\proof (proof of Theorem \ref{entrRateSpCase})

We begin the proof by noticing that $\forall k = \overline{1,K}$ $N_k\in (R^{K-k}-\alpha^{K-k}, R^{K-k}+\alpha^{K-k})$. Thus $\forall k = \overline{1,K}$ there are no more than $2\alpha^{K-k}$ possible integer values for $N_k$. Let $C(K,\alpha)$ be the total number of possible collections of Horton-Strahler numbers $N_1,N_2,\cdots, N_K$, such that $N_k = R^{K-k}\pm \alpha^{K-k}$. Notice that for every particular collection of Horton-Strahler numbers $N_1,N_2,\cdots, N_K$ there are
\begin{eqnarray*}
\label{e15}
|\mathscr{T}_{N_1,N_2,\cdots, N_K,R}| &=& 2^{N_1-1-\sum_{k=1}^{K-1}N_{k+1}}\prod_{k = 1}^{K-1} {N_k-2\choose 2N_{k+1}-2}
\end{eqnarray*}
planted binary plane trees. Thus, for a given set of parameters $K$ and $R$ the number of all planted binary plane trees with Horton-Strahler numbers $N_1,N_2,\cdots, N_K$ that satisfy $N_k = R^{K-k}\pm \alpha^{K-k}$, $\forall k = \overline{1,K}$ is given by
\begin{equation*}
|\mathscr{T}_{K,R}|= \sum_{(N_1,N_2,\cdots, N_K)\in C(K,\alpha)}|\mathscr{T}_{N_1,N_2,\cdots,N_K,R}|,
\end{equation*}
where $\mathscr{T}_{N_1,N_2,\cdots,N_K,R}\subset \mathscr{T}_{K,R}$.
Assuming uniform distribution of such trees, the probability of one tree $T_{K,R}\in \mathscr{T}_{K,R}$ is given by $P(T_{K,R}) = 1/|\mathscr{T}_{K,R}|$. Therefore, the entropy rate is given as
\begin{equation}
\label{fhlp0}
\mathscr{H}_{\infty}(K,R) = \lim_{N\rightarrow \infty}\frac{H(T_{K,R})}{N} = \lim_{N\rightarrow \infty}\frac{1}{N}\log_2|\mathscr{T}_{K,R}|.
\end{equation}
To find the entropy rate in (\ref{fhlp0}), first note that $|\mathscr{T}_{K,R}|$ can be bounded as follows
\begin{equation}
\label{hlp111}
|\mathscr{T}_{N_1^*,N_2^*,\cdots,N_K^*,R}|\leq |\mathscr{T}_{K,R}| \leq |\mathscr{T}_{N_1^*,N_2^*,\cdots,N_K^*,R}|\times C(K,\alpha),
\end{equation}
where $(N_1^*,N_2^*,\cdots,N_K^*) = \arg\max_{N_1,N_2,\cdots,N_K\in C(K,\alpha)} |\mathscr{T}_{N_1,N_2,\cdots,N_K,R}|$.
Since
\begin{equation*}
\label{totalTreesN}
C(K,\alpha) \leq \prod_{k=1}^{K}2\alpha^{K-k} = 2^K\alpha^{\sum_{k=1}^{K}(K-k)} = 2^K\alpha^{\frac{K(K-1)}{2}},
\end{equation*}
we can rewrite formula (\ref{hlp111}) in the following way
\begin{equation}
\label{trtr}
|\mathscr{T}_{N_1^*,N_2^*,\cdots,N_K^*,R}|\leq |\mathscr{T}_{K,R}| \leq |\mathscr{T}_{N_1^*,N_2^*,\cdots,N_K^*,R}|\times 2^K\alpha^{\frac{K(K-1)}{2}}.
\end{equation}
Next we apply the logarithm and divide by $N$ all sides of the inequality in (\ref{trtr}). Taking the limit as $N\rightarrow \infty$, we conclude that
\begin{equation*}
\label{e16}
\mathscr{H}_{\infty}(K,R) = \lim_{N \rightarrow \infty}\frac{\log_2|\mathscr{T}_{K,R}|}{N} = \lim_{N \rightarrow \infty}\frac{\log_2|\mathscr{T}_{N_1^*,N_2^*,\cdots,N_K^*,R}|}{N},
\end{equation*}
since
\begin{eqnarray*}
\lim_{N\rightarrow \infty}\frac{1}{N}\log_2\left(2^{K}\alpha^{\frac{K(K-1)}{2}}\right)&=& \lim_{N\rightarrow \infty}\frac{K+\frac{K(K-1)}{2}\log_2\alpha}{N} = 0.
\end{eqnarray*}
Finally, using results of Lemma \ref{lemma_hlp2}, we conclude that
$$\mathscr{H}_{\infty}(K,R) = 1-\frac{1-H\left(2/R\right)}{2-2/R}. $$
\endproof


\bibliography{poisson_noise}
\bibliographystyle{ieeetr}

\end{document}